# FPA Beamforming for Alignment-Tolerant FSO QKD Links

Florian Honz[(1)], Winfried Boxleitner[(1)], Michael Hentschel[(1)], Philip Walther[(2)], Hannes Hübel[(1)], and Bernhard Schrenk[(1)]

[(1)]AIT Austrian Institute of Technology, Center for Digital Safety & Security / Security & Communication Technologies, 1210 Vienna, Austria.
[(2)]University of Vienna, Faculty of Physics, Vienna Center for Quantum Science and Technology (VCQ), 1090 Vienna, Austria.
Author e-mail address: florian.honz@ait.ac.at

**Abstract:** We demonstrate focal plane array beamforming for semi-blind deployments of free-space optical QKD links. We accomplish a secure-key rate of 1.2 kb/s at a QBER of 9.1% over a 63-m out-door link during full sunshine.

## 1. Introduction

The rapid progress in quantum computer development threatens the security and longtime secrecy of data currently protected by asymmetric cryptography. Fortunately, quantum key distribution (QKD) offers a solution that can guarantee key exchange with information theoretic security, even against the possibilities of the quantum computer. Nowadays QKD systems have moved from research labs on to commercial systems being deployed in fiber networks around the globe. However, there are environments where fiber deployment is not feasible nor possible, necessitating the transition to free-space optical (FSO) communications. However, FSO-based QKD systems have not reached the same maturity as their fiber-based counterparts yet, mainly due to size and complexity, since large telescopes are needed to ensure good coupling efficiency and active alignment of the single mode fiber (SMF) based transmitter and receiver stations is paramount [1-4]. While recent demonstrations have successfully shown the adoption of photonics to reduce the footprint of free-space QKD setups [1] as well as the use of large-core optics on the receiver side to enhance the alignment tolerance [5], they either necessitate complex active beam steering in the case of single mode to single mode coupling or are reach-limited due to the employed multi-mode fibers on the receiver side.

In this work we adopt focal plane array (FPA) beamformers to enhance the alignment tolerance of a free-space link based on SMF-to-SMF coupling. In addition, we use a compact photonic-integrated QKD state preparation circuit and show successful co-propagation of the QKD signal with 47×10 Gb/s classical data channels at an aggregated power of 11.2 dBm, reaching a secure-key rate of 785 b/s. We further demonstrate QKD over a 63-m out-door bridge.

## 2. Focal Plane Array Beamforming for Enhanced Alignment Tolerance of Free-Space Optical Transmission

Our point-to-point FSO link is realized by two FPA beamformers as fiber-to-air interfaces for two ITU-T G.652B-compatible SMFs at the transmitter and receiver sides (Fig. 1a). Each FPA consists of 61 individual SMF cores of a photonic lantern with a mode-field diameter of 8.4 µm and a pitch of $\delta$ = 36.9 µm, arranged in a hexagonal pattern with up to $N$ = 9 elements per axis. In combination with a suitable collimation lens, the direction of the emitted pencil beam can be changed as a function of the offset of the light feed from the focal center of the lens [6], which is facilitated through a simple switching of the optical antenna element (i.e., focal-plane core). The fill factor of all cores over the focal plane is 5.4% or -12.7 dB. Photonic lanterns provide access to each antenna element through individual SMF ports. The 2" collimation lens completing the FPA assembly has a focus of $f$ = 150 mm, leading to beam diameter of ~32 mm and a field-of view of $FoV = N \cdot \delta / f$ = 0.13° (14 cm at a link distance of $L$ = 63 m).

By using a 1×31 optical switch (SWI) at the transmitter side and a 61×1 SWI at the receiver side, the antenna elements at the transmitter and receiver terminals can be selected according to a periodic and time-interleaved classical channel sounding procedure prior to key exchange. The given FPA beamformer allows for a flexible switching of the beam emission angle and thereby an enhanced field-of-view compared to direct SMF-to-SMF coupling, without involving complex control. This enables SMF coupling even under unfavorable initial alignment conditions and offers an improved robustness to external perturbations, as proven recently for classical Fi-Wi-Fi deployment scenarios [7].

## 3. QKD State Preparation on Silicon Photonics and Experimental Setup of Hybrid Quantum/Classical Link

Figure 1b presents our experimental setup, employing a polarization-based BB84 DV-QKD protocol. Our QKD state preparation in the C-band at $\lambda_Q$ = 1550.12 nm operates at a symbol rate of 1 GHz and is hosted on a 2.5 mm² photonic integrated circuit (PIC) based on silicon-on-insulator technology. After coupling light through a 1D grating coupler (GC), the first phase modulation section PM1 together with a multi-mode interference (MMI) coupler is flexibly partitioning the optical power between the two branches of the second phase shifter PM2. Since the 2D-GC at the PIC output performs a conversion from the TE- to the TM-mode for one of its branches, each branch of PM2 corresponds to one of the two principal polarization states (**H**, **V**). In combination with PM2, which shifts the relative phase between the **H** and **V** components, any arbitrary polarization state can be generated. However, in our implementation of the BB84 protocol we resort to the four states **H**, **V**, **R** and **L**. We additionally employed an off-chip Mach-Zehnder-Modulator (MZM) for (*i*) super-Gaussian shaped pulse-carving to 50% of the symbol width and (*ii*) decoy state transmission. Despite the high fiber-to-fiber PIC loss of 28 dB, which is primarily attributed to its long phase-shifters, we still require an external variable optical attenuator (VOA$_Q$) to set the desired launch level of $\mu$ = 0.1 $h\nu$/symbol.

The quantum signal is then multiplexed with C+L+U band classical signals through a wideband 50-GHz DWDM add/drop (A/D) filter centered on $\lambda_Q$. The 48 classical channels spanned from 1530.25 to 1618.63 nm and are located in-band to the C-band quantum channel. The 47 data channels were on-off keyed at 10 Gb/s (Fig. 2a) and boosted by C- and L-band EDFAs while the U-band channel 48 served as a supervisory channel. A quadruple notch filter $T_Q$ is employed to eliminate any noise photons at $\lambda_Q$ by virtue of its high suppression of 132.3 dB at 1550.12 nm (Fig. 2a).

After free-space transmission over the two FPA-enhanced air interfaces, the quantum and classical signals are demultiplexed by an identical DWDM A/D filter, before quantum state analysis is performed with a fiber-based polarization analyzer (PA). A manual polarization controller is used to align the polarization with the axis of the PA and to compensate for polarization drift during signal transmission. Each output of the PA then corresponds to one state of the non-orthogonal **R/L** and **H/V** basis and detection is performed by a pair of InGaAs SPADs with a detection efficiency of 10%, a dead-time of 25 µs and dark count rates of 559 cts/s and 599 cts/s, respectively. The detection events of the SPADs are recorded by a TTM with a resolution of 82.3 ps, before an off-line evaluation is performed to calculate and estimate the raw- and secure-key rates as well the quantum bit error ratio (QBER).

## 4. QKD Performance over FPA-Enhanced FSO Bridge

We first set up the FPA transmitter and receiver terminals for a FSO link with a distance of $L$ = 6 m within our lab, which features an average background irradiance of ~800 lux due to ambient sunlight. The terminals are set by simply performing a coarse visual alignment with red beacon lasers. We then connect the classical signal feed to perform an automated alignment of the FSO bridge, for which the beamforming at the receiver-side terminal is remotely controlled from the transmitter side through an auxiliary wireless control channel. The calibration map for the FSO bridge is reported in Fig. 2b in terms of optical coupling between pairs of FPA antenna elements $\zeta_{TX}$ and $\xi_{RX}$. Following the semi-blind pointing of the FSO terminals, we achieve a minimum FSO loss of 15.5 dB between $\zeta_{TX}$ = 49 ($\varepsilon$ in Fig. 2b) and $\xi_{RX}$ = 33, including 2 dB of insertion loss per FPA for the employed MEMS switches and photonic lanterns. The offset from the ideal center-center coupling results from the coarse initial alignment. Figure 2c shows the coupled power for all receiver-side FPA antenna elements for this optimal launch at $\zeta_{TX}$ = 49. We couple mainly into a single element ($\xi_{RX}$ = 33), given that the optical power coupled to the second-best ($\xi_{RX}$ = 32) already drops by 14.4 dB.

We first evaluated the QKD system over a 6-m long SMF span to determine a "back-to-back" performance baseline (● in Fig. 3a). For an optical budget of 0 dB, meaning that QKD transmitter and receiver are interfaced without extra link loss, we reach a raw-key rate of 54.3 kb/s at a QBER of 2.07%. The QBER limit of 11% for secure-key extraction is surpassed for an optical budget of 26 dB. This would correspond to an end-of-life fiber loss of 94 km of SMF (0.277 dB/km). When switching QKD transmission to the FSO link (♦), we reach a raw-key rate of 23.5 kb/s at a QBER of 4.9% for $\mu$ = 0.1 photons/symbol. There is no additional penalty induced apart from the link loss of 15.5 dB over the FSO bridge, which erodes part of the raw-key rate. An extra optical budget of 10.5 dB can be accommodated before reaching the QBER threshold. Comparison with the calibration map of the FSO link (Fig. 2b) shows that there are 28 FPA pairs that offer a sufficiently high margin to enable secure-key generation, with five of those having a power penalty of less than 3 dB. This diversity in FPA pairs can be exploited upon a fading FSO channel in order to re-establish better channel conditions through alternative antenna elements $\zeta_{TX}$, $\xi_{RX}$.

In a next step we loaded the FSO link with all 48 classical channels adjacent to $\lambda_Q$. There was no additional QKD performance penalty under this co-existence condition as long as the aggregated classical power level $P_{Cla}$ remained below 0 dBm (Fig. 3b). For an elevated launch power we noticed the presence of in-band Raman noise at the receiver, despite the exclusive use of short fiber spans of less than 100 m. This is attributed to the spectral classical/quantum co-existence setup where (in-band) data channels continuously populate a wide spectral detuning range of 155.5 GHz to 8.19 THz from the quantum channel. However, we can still accomplish a raw-key rate of 17.5 kb/s at a QBER of 10.3% for a classical launch power of 11.2 dBm. This yields a secure-key rate of 785 b/s, which allows to secure up to 1.57 Tb/s of classical data capacity under the NIST limit for AES-GCM key-renewal (i.e., one 256-bit key to be consumed per 64 GB of data). It is therefore sufficient to secure the full FSO link capacity of 470 Gb/s.

The BER performance of representative data channels in the lower/upper C/L-band is shown in Fig. 3c as function of the received optical power (ROP). Apart from the band-specific gain of the preamplified EDFA+PIN receiver, which show a gain excursion especially in the C-band due to omission of gain flattening filters, data transmission can be established for all channels below the forward error correction (FEC) limit of $10^{-3}$.

Finally, we tested the QKD performance over a 63-m out-door link (Fig. 2d) with a loss of ~20 dB. Despite narrow optical filtering, the in-band noise due to the high solar irradiance of ~61 klux led to increased background counts of 1204 cts/s and 980 cts/s, respectively. Nonetheless, we could achieve a QBER of 9.13% at a raw-key rate of 10.1 kb/s (Fig. 3e), yielding a secure-key rate of 1.2 kb/s. Longterm operation would require active control for the beamforming.

## 5. Conclusion

We have demonstrated a FSO QKD link with increased alignment tolerance by using a simplified FPA beamformer for the purpose of SMF-to-SMF coupling. In addition, we are able to simultaneously secure 47 classical channels co-propagated with the quantum signal over the same link and demonstrated secure-key generation under strong sunlight. A further reduction of the link loss is feasible through an improvement of the current FPA fill factor of -12.7 dB.

**Acknowledgement**: This work was supported in part by the European Research Council (ERC) under the EU's H2020 programme (GA No 804769), through the QuantERA II Programme (GA No 101017733) and with the Austrian FFG (GA No FO999906040).

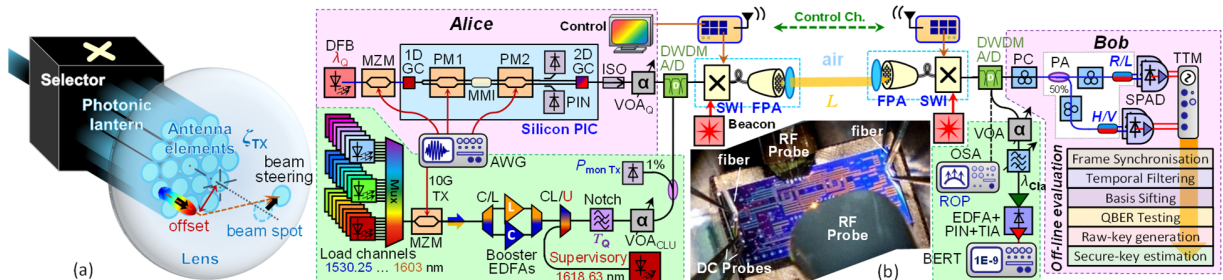

Fig. 1. (a) FPA beamformer. (b) Experimental setup for FPA-enhanced QKD/classical FSO link. Inset: PIC-based quantum state preparation.

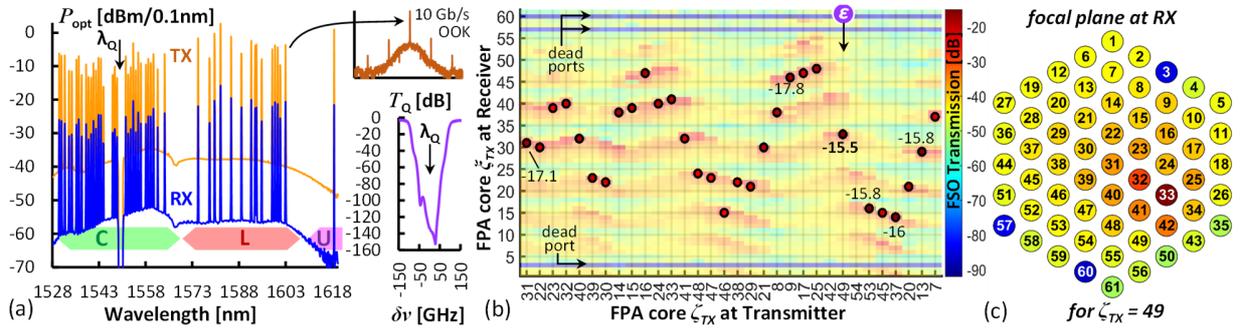

Fig. 2. (a) Spectral allocation of the quantum / classical 10 Gb/s channels and notch filter $T_Q$ to clear classical noise. (b) Coupling between optical antenna elements at FSO transmitter and receiver. (c) Received optical power per RX antenna element for the optimally coupled TX channel.

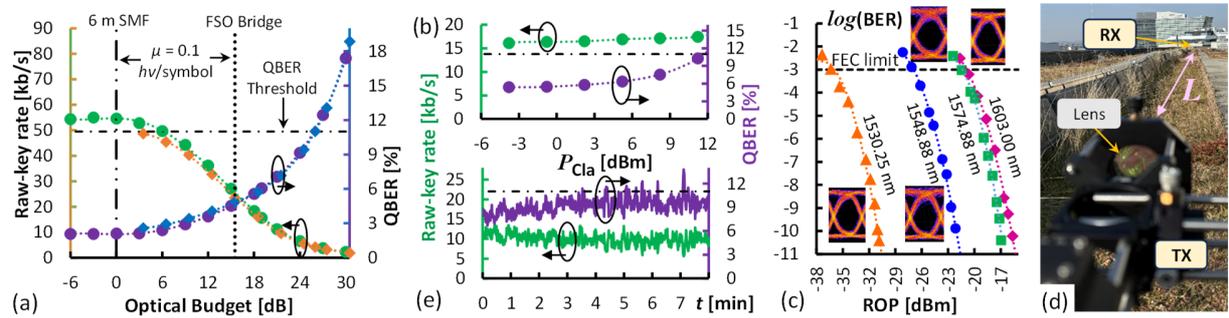

Fig. 3. (a) QKD performance over back-to-back (●) and over FSO channel (◆). (b) QKD performance under classical co-existence. (c) BER performance of four exemplary classical channels. (d) Out-door FSO setup and (e) QKD performance over out-door FSO link.